\documentclass[aps,prl,twocolumn,showkeys,showpacs,amsmath,amssymb]{revtex4}

\usepackage[dvips]{graphicx}

\begin{document}

\title{Granular Avalanches in Fluids}

\author{Sylvain Courrech du Pont$^{1}$}
\author{Philippe Gondret$^{1}$}
\author{Bernard Perrin$^{2}$}
\author{Marc Rabaud$^{1}$}
\affiliation{$^{1}$Laboratoire Fluides, Automatique, Systemes
Thermiques (UMR CNRS 7608),\\
B\^at. 502, Campus Universitaire, 91405 Orsay cedex, France}
\affiliation{$^{2}$Laboratoire de Physique de la Matiere Condens\'ee (UMR
CNRS 8551),\\
24, rue Lhomond, 75231 Paris cedex, France}

\date{\today}

\begin{abstract}
Three regimes of granular avalanches in fluids are put in light
depending on the Stokes number St which prescribes the relative
importance of grain inertia and fluid viscous effects, and on
the grain/fluid density ratio r. In gas ($r \gg$ 1 and St $>$ 1,
e.g., the dry case), the amplitude and time duration of
avalanches do not depend on any fluid effect. In liquids
($r \sim$ 1), for decreasing St, the amplitude decreases and
the time duration increases, exploring an inertial regime
and a viscous regime. These regimes are described by the
analysis of the elementary motion of one grain.
\end{abstract}



\maketitle

Granular matter has received much attention from physicists over the
past few years \cite{Duran2000}. Beyond the fundamental interest in the
physics of
granular systems which can present some features of either solids,
liquids or even gases, the understanding of granular materials is
essential in many industrial activities such as pharmacology, chemical
engineering, food, agriculture, and so on. Many studies concern the
avalanches that may arise on the slope of a granular pile in air.
Such granular avalanches occur in various places in Nature, from
small scale, as for the building of any sand pile, to large scale,
as the event observed after the Mont St-Helen eruption in 1980.
Two angles can be defined when building a pile: the maximum angle
of stability $\theta_{m}$ at which an avalanche starts and the angle
of repose
$\theta_{r}$ at which the avalanche stops. Between these two angles is
a region
of bistability where the grains can either be flowing (``liquid state")
or at rest (``solid state"). Many experiments performed with dry grains
in a rotating cylinder
\cite{EveRajch88,Jaeger89,Rajchen90,Evesque91,Fauve94} showed clearly
the existence of these two angles.

To date, no detailed study has focused on the influence of the interstitial
fluid for a totally immersed grain assembly. This influence is certainly
important in granular avalanche processes, as evidenced by the marked
differences observed by geologists between subaqueous and eolian cross
strata \cite{Hunter85}. As a matter of fact, the propagation of subaqueous
dunes
differs in general from the propagation of eolian dunes even if the slope
angles are quite similar: When the transport rate of sand particles is
large enough, the flow is continuous in the lee side of the structure
in the immersed case, but occurs by successive avalanches in the dry
case \cite{Hunter85}. This observation prompted geologists to accumulate
data on
avalanches of sand or beads in rotating drums filled with air or
water \cite{Carrigy70} or even with glycerol mixtures \cite{Allen70a}, that
seemed to show
that the amplitude of avalanches decreases and the time duration
increases with the fluid viscosity. We have performed an extensive
series of experiments to investigate the influence of the interstitial
fluid on the packing stability and the avalanche dynamics. The
analysis of our results obtained with a rotating drum set-up indicate
the existence of three regimes: (i) a free-fall regime for which there
is no fluid influence and that corresponds to the classical dry regime,
and two regimes where the interstitial fluid governs the avalanche
dynamics, namely (ii) a viscous regime and (iii) an inertial regime.

Our rotating drum consists of a cylinder of inner diameter $D$ ranging
from 8 cm to 46 cm and lying on two parallel rotation axes.
It is driven by a microstep motor followed by a 1/100 reducer
and a rubber transmission so that the cylinder turns at the rotation
rate $\Omega$ by step of 10$^{-3}$ degree without shocks. The cylinder is half
filled with sieved solid spheres of diameter $d$ and density $\rho_{s}$,
totally
immersed in a fluid of density $\rho_{f}$ and dynamic viscosity $\mu$.  The
pile
is confined between two parallel glass endwalls separated by the gap
width $b$. In each experimental configuration, we used a sufficiently
large gap width ($b/d$ $>$ 15). This study was achieved at
low enough rotation rate $\Omega$ to be in the intermittent regime of
macroscopic avalanches \cite{Rajchen90,Fauve94}: The pile slope increases
linearly with
time at the rate $\Omega$, and then quickly relaxes by a surface avalanche
process. A CCD camera aligned along the axis of the cylinder
allows for visualization of the rotating pile. Images are taken at
regular time intervals and then analyzed to track the
pile interface. Except during
the avalanches, the interface is found to be linear, with a roughness
of the order of one grain size, which justifies the calculation of
the averaged slope angle $\theta$  of the pile with the resolution
$0.01^\mathrm{o}$.

\begin{figure}
\includegraphics[width=8cm]{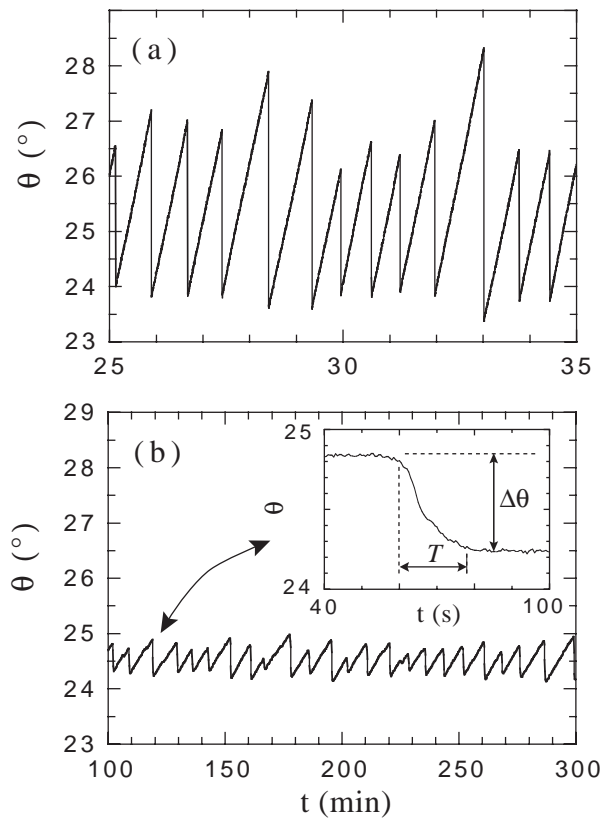}\\
\caption{Time evolution of the average slope
angle $\theta$ of a pile of glass beads ($d$ = 230 $\pm$ 30
$\mu$m)
immersed in air (a) or water (b) and contained in a rotating
drum. The insert is a zoom of an avalanche in the water case.}
\end{figure}

Two typical recordings for the time evolution of the average slope angle
$\theta$ are displayed in Fig. 1 for 230 $\mu$m glass beads immersed either
in air
or water. We focus on two typical parameters of the avalanche dynamics
that we found uncorrelated: The avalanche amplitude characterized by
the hysteresis angle $\Delta \theta$ = $\theta_{m}$ - $\theta_{r}$,
and the avalanche time duration $T$
which is calculated as the time interval between 5$\%$ and 95$\%$ of the
corresponding avalanche amplitude. The mean values $\overline{\Delta
\theta}$ and $\overline{T}$  are
then calculated for one experiment over all the successive
macroscopic avalanches that affect the entire slope. In the following,
we will drop the mean bar notations for simplicity. Despite the dispersion,
two different behaviors clearly appear:  In the air case (Fig. 1a), the
avalanche amplitude $\Delta \theta$ is large (few degrees) and the
avalanche duration
$T$ is small (typically one second), whereas in the water case (Fig. 1b)
$\Delta \theta$ is small (less than one degree) and $T$ is large (typically
one minute).
One crucial parameter of the phenomenon appears to be the particle
diameter $d$ when one looks in Fig. 2 at the evolution of $T$ and $\Delta
\theta$ as a
function of $d$ when the grains are immersed either in air or in water.
In the air case, there is no significant dependency of $T$ and $\Delta
\theta$ on $d$.
By contrast, in the water case, the avalanche duration which is close to
the air case for the larger $d$, increases first sligthly ($T \propto
d^{1/2}$) then
drastically ($T \propto d^{2}$) from typically 2 to 100 s when the grain
size is
decreased from 1 mm to 0.18 mm. Parallely, the avalanche amplitude $\Delta
\theta$
in water corresponds to the air value for the larger $d$ but decreases
close to zero when the grain size is decreased to 0.18 mm.

\begin{figure}
\includegraphics[width=7cm]{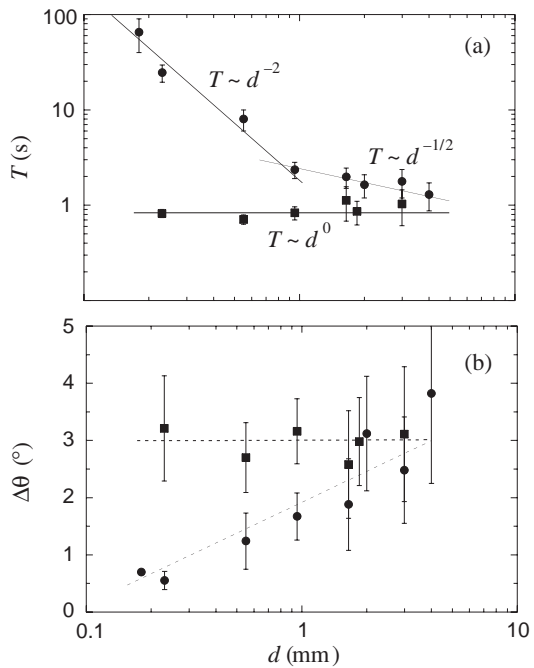}\\
\caption{Time duration $T$ and amplitude $\Delta \theta$ of
macroscopic avalanches as a function of the grain diameter
$d$ for glass beads immersed in air ($\blacksquare$) or water
($\bullet$) in a rotating cylinder of diameter $D$ = 16 cm.
The error bars correspond to the standard deviation.}
\end{figure}

The pertinent dimensionless parameters governing the avalanche dynamics
in fluids can be inferred by considering the elementary falling process
of one solid grain on its neighbour from below in a fluid and under the
action of gravity. Let us write the simplified following equation of
motion for the grain of velocity $u$ down the slope between two collisions:
\begin{equation}
    {\pi \over 6} \rho_{s} d^{3} {du \over dt} =
    {\pi \over 6} \Delta \rho g d^{3} \mathrm{sin}\theta - F_{d}.
\end{equation}

Starting from zero velocity, the grain increases its grain momentum at
the rate $(\pi / 6) \rho_{s} d^{3} (du/dt)$ under the action of its
apparent weigth
$(\pi / 6) \Delta \rho g d^{3} \mathrm{sin}\theta$ (where
$\Delta \rho$ = $\rho_{s}$ - $\rho_{f}$) minus a fluid drag force $F_{d}$.
At this basic
stage, the solid friction force can be modelled by a dynamical
Coulombic term which just reduces the apparent gravity. Two behaviors
can be discussed according  to the value of the particle Reynolds number:
At Re $\ll$ 1, $F_{d}$ is equal to the viscous Stokes force $3 \pi d
\eta u$  so the grain will
possibly reach its  viscous limit velocity  $U_{\infty v}$ = $\Delta
\rho g d^{2} \mathrm{sin}\theta/18 \eta$   in the characteristic
time $\tau_{cv}$  = $\rho_{s} d^{2}/18 \eta$, i.e. for a
characteristic
distance  $\delta_{cv}$ = $\tau_{cv} U_{\infty v}$. At Re $\gg$ 1,
$F_{d}$ is the inertial fluid force $C_{d} (\pi / 6) d^{2} \rho_{f} u^{2}$,
and the inertial characteristic time and distance are
$\tau_{ci}$ = $(\rho_{s}/\rho_{f})^{1/2}(2 \rho_{s}d/\Delta \rho g
\mathrm{sin}\theta)^{1/2}$ and  $\delta_{ci}$ =  $\tau_{ci} U_{\infty i}$
where  $U_{\infty i}$ = $(2\Delta \rho g d \mathrm{sin}\theta/
\rho_{f})^{1/2}$ is the
inertial limit velocity (for simplicity we take here the drag
coefficient as constant : $C_{d} \simeq 1/\pi \simeq 0.3$). By comparing
the two
characteristic distances $\delta_{cv}$ and $\delta_{ci}$ with the
elementary distance
beween two successive collisions taken as the grain diameter $d$,
we introduce two dimensionless numbers, St and $r$, which govern
the grain dynamics in this elementary falling process:
$\delta_{cv}/d$ = $2(\tau_{cv}/\tau_{ff})^{2}$ = $2 \mathrm{St}^{2}$ and
$\delta_{ci}/d$ = $2(\tau_{ci}/\tau_{ff})^{2}$  = $2 r^{2}$
where $\tau_{ff}$ = $(2 \rho_{s}d / \Delta \rho g
\mathrm{sin}\theta)^{1/2}$ is
the typical timescale of free falling of a grain over $d$,
St = $(1/18\sqrt{2})\rho_{s}^{1/2}(\Delta \rho g
\mathrm{sin}\theta)^{1/2}d^{3/2}/\eta$
is the Stokes number which prescribes the relative
importance of grain inertia and fluid viscous effects of
and $r$ = $(\rho_{s}/\rho_{f})^{1/2}$ is related to the density ratio. Note
that the Reynolds number corresponds to the third time ratio,
so that Re = $\tau_{cv}/\tau_{ci}$ = St/$r$. For St $\gg$ 1 and $r \gg$ 1,
the grain does
not reach any limit regime: This is the ``free-fall regime". For
St $\ll$ 1 and $r \gg$ 1 the grain reaches its limiting Stokes velocity:
This is the ``viscous limit regime". For St $\gg$ 1 and $r \ll$ 1, the
grain reaches its limiting inertial velocity: this is the ``inertial limit
regime". For St $\ll$ 1 and $r \ll$ 1, the grain
reaches one limit velocity depending on the Re = St/$r$ value:
for Re $\ll$ 1 (resp. Re $\gg$ 1) the limit regime is the viscous
(resp. inertial) one. The exact boundaries between the three
domains in the (St, $r$) plane of Fig. 3 will be precised further.
In this diagram are reported all our data and other data
\cite{Evesque91,Allen70a},
corresponding to different sphere materials (glass and Nylon)
in different fluids (air, water, silicone oils or glycerol
mixtures of different viscosities). All experimental results
correspond to roughly two data lines in this diagram: one data
line for the liquid case where $r \sim 1$ and St ranges from 0.2 to 40
and another data line for the air case where $r \sim 40$ and St ranges
from 30 to $10^{4}$.

Let us now look if the complex dynamics of macroscopic granular
avalanches in fluids can be related to these elementary falling
processes. In the two limit regimes, one may reasonably suppose
that the time duration $T$ of a macroscopic granular avalanche will
scale as $D/d$ elementary fallings each of time duration $d/U_{\infty v}$, so
that $T_{v}$ = $D/U_{\infty v}$  = $18 \eta D / \Delta \rho g d^{2}
\mathrm{sin}\theta$ in the viscous limit regime and $T_{i}$ =
$D/U_{\infty i}$  = $D \rho_{f}^{1/2}/(2\Delta \rho g d
\mathrm{sin}\theta)^{1/2}$ in the
inertial limit regime. By the way, the two scalings $T \propto d^{-2}$ and
$T \propto d^{-1/2}$  observed in Fig. 2 for small and large $d$ respectively
correspond to these two predicted scalings. We have plotted $T/T_{v}$
and $T/T_{i}$ in Fig. 4a and 4b respectively and we observe that all
data collapse onto the plateau $T \simeq 4 T_{v}$ for St $\lesssim$ 5 (Fig
4a),
and the plateau $T \simeq 2 T_{i}$ for St $\gtrsim $ 3 (Fig 4b). A careful
analysis
of each data point of Fig 4a,b allow us to identify the critical
Reynolds number $\mathrm{Re}_{c} \simeq $ 2.5 for the viscous/inertial
transition,
and thus to draw the corresponding boundary line of slope 1
between the two corresponding domains in the log-log plane
(St, $r$) of Fig. 3.

\begin{figure}
\includegraphics[width=7cm]{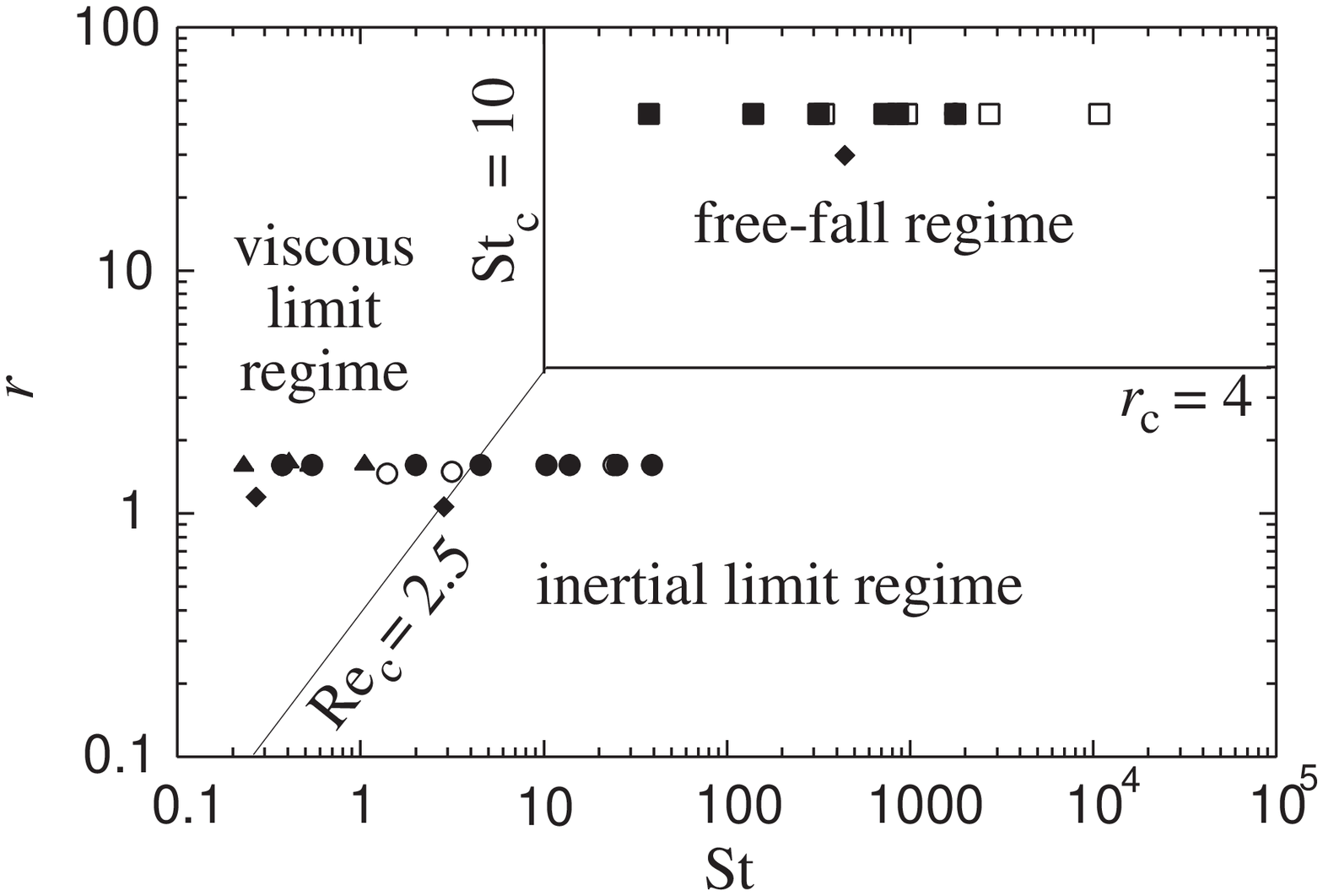}\\
\caption{Diagram of the different regimes for the elementary falling
process of one grain in the (St, $r$) plane.  The three regimes
are separated by the three boundaries $\mathrm{St}_{c}$ = 10,
$\mathrm{Re}_{c}$ = 2.5 and
$r_{c}$ = 4. Our results (filled symbols) correspond to glass beads
in air ($\blacksquare$) or water ($\bullet$), or in silicone oils of different
viscosities ($\blacktriangle$), and to Nylon spheres ($\blacklozenge$) in
air, water and
silicone oil. Other results (open symbols) of Evesque [5] ($\square$)
and Allen [10] ($\circ$) correspond to glass beads in air and
different water-glycerol mixtures.}
\end{figure}

In the elementary free-fall regime corresponding to the air
case (St $\gg$ 1 and $r \gg$ 1), the same approach would lead to the
avalanche time duration $T$ = $(D/d) \tau_{ff}$ = $D (2 \rho_{s} / \Delta
\rho g d \mathrm{sin}\theta)^{1/2}$, which is not consistent
with the non $d$ dependence observed in Fig. 2. In addition, we have
also observed that $T$ scales as $D^{1/2}$ in the air case rather than as
$D$. All these observations lead us to consider a dry avalanche as a
global accelerated rush over $D$ of macroscopic timescale $T_{ff}$ = $(2
\rho_{s}D / \Delta \rho g \mathrm{sin}\theta)^{1/2}$
rather than a succession of elementary falling processes. This
scaling is indeed observed as all data in air collapse onto the
plateau $T \simeq 3 T_{ff}$  for St $\gtrsim$ 30 (Fig. 4c). In all three
regimes,
the plateau values of Fig. 4a,b,c are not far from one, meaning
that our crude approach catches the essential of the avalanche
dynamics in fluids. Note that for much larger $D$ inaccessible to
laboratory experiments, such a free-fall regime will possibly not
be observed as the inertial fluid force or a solid friction force
will come into play at large grain velocity.

\begin{figure}
\includegraphics[width=8cm]{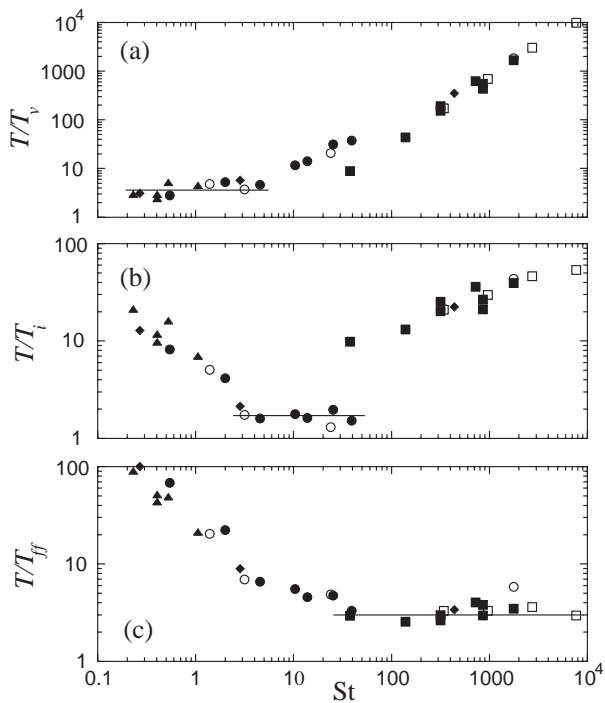}\\
\caption{Time duration $T$ of macroscopic avalanches normalized either by
(a) the viscous timescale $T_{v}$ (b) the inertial timescale $T_{i}$ or
(c) the free-fall timescale $T_{ff}$ as a function of the Stokes
number St for different grains in different fluids and rotating
cylinders of different diameter $D$. Same symbols as in Fig. 3.}
\end{figure}

Let us focus now on the avalanche amplitude $\Delta \theta$.
When plotting $\Delta \theta$  as a function of the Stokes number
St (Fig. 5), we obtain a single master curve. At large St
(St $\gtrsim$ 20) $\Delta \theta$  has the constant plateau value
$\Delta \theta \simeq 3^\mathrm{o} \pm 1^\mathrm{o}$
whereas $\Delta \theta$ clearly decreases with St at small St (St
$\lesssim$ 20).
We also observe that the decrease of $\Delta \theta$  at small St is
essentially due to a decrease of the maximum angle of
stability $\theta_{m}$, the angle of repose $\theta_{r}$ being almost
unchanged
(cf. Fig. 1). This can be explained by considering the
collision process beween immersed grains. Indeed, the
Stokes number is the only parameter that governs the
coefficient of restitution $e$ for the immersed binary
collision between solid grains \cite{Gondret2002}. Under the critical
value $St_{c} \simeq$ 10, the coefficient of restitution is zero:
the collision is totally inelastic as all the kinetic
energy of the grain is dissipated by the fluid during
the collision process \cite{Gondret2002}. Above $St_{c}$, the coefficient
of restitution increases quickly with St and becomes
close to its maximal ``dry" value $e_{dry} \simeq$ 1 for St $\gtrsim$ 100
\cite{Gondret2002}.
It is worthnoting that the curve $e$ = f(St) is the same
whatever the density ratio $r$ \cite{Gondret2002}. Considering again the
immersed granular avalanches in fluids, the grain kinetic
energy will be totally dissipated by the fluid in the
collision process at low St, with an all the more smooth
collision when St evolves towards zero. The obtained packing
is thus certainly all the more loose, i.e. with a lower
packing fraction \cite{Onoda90}. As the maximum angle of stability of
a granular pile depends largely on the arrangement of the
packing, decreasing with the packing fraction \cite{Allen70b}, this
explains the decrease of $\Delta \theta$  with St at low St.

In addition, if one consider the critical Stokes value
$\mathrm{St}_{c} \simeq$ 10 independent of $r$ as the boundary line between
the accelerated regime and viscous limit regime in the
(St, $r$) plane of Fig. 3, this leads to the critical
density ratio $r_{c} \simeq$ 4 separating the free-fall regime
and the inertial limit regime. As $r$ values larger than
$4(\rho_{s}/\rho_{f} >$ 16) can hardly be reached experimentally for
solid/liquid system, the free-fall regime corresponds only
to solid/gas systems like, e.g., the dry granular avalanches.

Finally, for large St (St $\gtrsim$ 20), all the events correspond
to macroscopic avalanches that affect the entire slope and
no small event are observed between two successive macroscopic
avalanches: The size distribution is a gaussian curve centered
on the value $\Delta \theta \simeq 3^\mathrm{o}$. This kind of
distribution, classically
found for dry granular avalanches
\cite{EveRajch88,Jaeger89,Rajchen90,Evesque91,Fauve94}, is incommensurate with
the ideas of Self Organized Criticality (SOC) developed by Bak
et al. \cite{Bak87} which would predict a power law distribution without
any typical scale. The reason may be the dissipation rate of the
system [16]: In the cellular automata models, which illustrate
nicely the SOC, automata are strongly overdamped whereas dry granular
experiments are weakly dissipative. When introducing inertia in cellular
automata models, a complex distribution, mixing power law distribution
for the small events and gaussian distribution for the large events,
is obtained \cite{Prado92}. For decreasing St (St $\lesssim$ 20), we
observe experimentally
together with the gaussian distribution of large events the appearance
of numerous small events (affecting not all the slope). We expect such
a behavior in the viscous regime of low St as the hysteresis $\Delta
\theta$  of the
system goes to zero with St, which is a condition for the system to
evolve towards criticality. But up to now, we have not enough resolution
to characterize the size distribution of these small events, and to
conclude if it obeys or not the power law related to SOC. In addition,
the regime of intermittent avalanches is hard to obtain when the regime
is more and more viscous as the time duration of avalanches diverges.

\begin{figure}
\includegraphics[width=7cm]{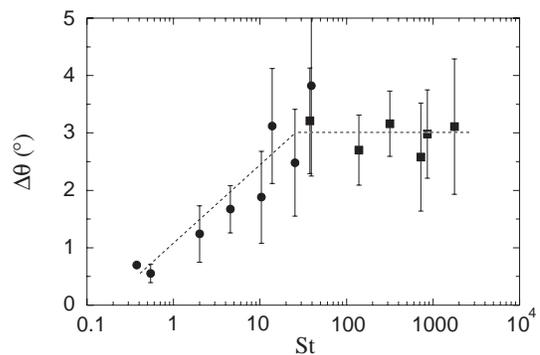}\\
\caption{Amplitude $\Delta \theta$ of macroscopic avalanches of glass beads
in different fluids as a function of the Stokes number St.
Same symbols as in Fig 3.}
\end{figure}

By conclusion, we have shown the existence of three regimes
(free-fall, inertial limit, and viscous limit) for granular
avalanches in fluids, controlled by the Stokes number which
measures the ratio of particle inertia to viscous fluid effects,
and the density ratio. The time duration of the macroscopic
avalanches that affect the entire slope have been predicted in
all these three regimes. The amplitude of these macroscopic
avalanches has been shown to be constant at high St while
decreasing with St at low St. Finally, more refined experiments
remain to be done to see if the system evolves towards criticality
when St tends towards 0, i.e. for highly dissipative systems.

We acknowledge B. Andreotti, S. Douady, D. Lhuillier, O. Pouliquen,
E.J. Hinch, and G.M. Homsy for fruitful discussions.

\end{document}